\documentclass[aps,preprint,showpacs, showkeys]{revtex4}%
\usepackage{amsfonts}
\usepackage{amsmath}
\usepackage{amssymb}
\usepackage{graphicx}%
\providecommand{\U}[1]{\protect\rule{.1in}{.1in}}

\begin{document}
\title[ ]{A Novel phase in the phase structure of the ($g\phi^{4}+h\phi^{6}$)$_{1+1}$
field theoretic model}
\author{Abouzeid. M. Shalaby\footnote{E-mail:amshalab@ mans.eg.edu}}
\affiliation{Centre for Theoretical Physics, The British University in Egypt, El Sherouk
City, Misr Ismalia Desert Road, Postal No. 11837, P.O. Box 43, Egypt }
\affiliation{Physics Department, Faculty of Science, Mansoura University, Egypt.}

\begin{abstract}
In view of the newly discovered  and physically acceptable $PT$ symmetric and non-Hermitian models, we reinvestigated the phase structure of the ($g\phi^{4}+h\phi^{6}$)$_{1+1}$ Hermitian model. The reinvestigation concerns the possibility of a phase transition from the original Hermitian and $PT$ symmetric phase to a non-Hermitian and $PT$ symmetric one. This kind of phase transition, if verified experimentally, will lead to the first proof that non-Hermitian and $PT$ symmetric models are not just a mathematical research framework but are a nature desire.   To do the investigation, we calculated the effective potential up to second order in the couplings and  found a Hermitian to Non-Hermitian phase transition. This leads us to introduce, for the first time, hermiticity as a symmetry which can be broken due to quantum corrections, \textit{ i.e.}, when starting with a model which is Hermitian in the classical level, quantum corrections can break hermiticity while the theory stays physically acceptable. In fact, ignoring this phase will lead to violation of
universality when comparing this model predictions with other models in the same class of universality. For instance, in a previous work we obtained a second order phase transition for the $PT$ symmetric and non-Hermitian $(-g\phi^{4})$ and according to universality, this phase should exist in the phase structure of the ($g\phi^{4}+h\phi^{6}$) model for negative $g$.  Finally, among the novelties in this letter, in our calculation for the effective potential, we introduced a new renormalization group equation which describes the invariance of the bare vacuum energy under the change of the scale. We showed that without this invariance, the original theory and the effective one are inequivalent. 

\end{abstract}
\maketitle

In a previous work we investigated the broken symmetry phase of the $PT$
symmetric and non-Hermitian $\left(  -g\phi^{4}\right)  $ quantum field model \cite{abphi4}.
We found a second order phase transition with a zero critical coupling in the
sense that both the mass parameter and the vacuum condensate goes to zero as
$g\rightarrow0$.  Such kind of phase transition may be helpful in describing the so called quantum phase transitions (at zero temperature) \cite{2dqm}. Moreover, with a $\phi^6$ term, as we will explain later in this letter, this kind of phase transition may be helpful in simulating second order transition out of a collinear Neel phase to a
paramagnetic spin liquid in two dimensional quantum antiferromagnets.
\cite{antiferro}.

Near a second order phase transition, it is well known that the $\phi^{6}$
operator in the $\left(  g\phi^{4}+h\phi^{6}\right)  $ quantum field model
(Hermitian) is irrelevant. Accordingly, the Hermitian $\left(  -g\phi
^{4}+h\phi^{6}\right)  $ model has to show up the same phase discovered for 
$\left(  -g\phi^{4}\right)  $ theory. However, in view of all of the previous investigations
of the $\left(  g\phi^{4}+h\phi^{6}\right)  $ model \cite{montecarlo,stft,gep6,orpap}, no such phase has been
discovered which   wrongly leads to     violation of universality.  On the other
hand, it has been  shown that the double exchange Hamiltonian, with weak
antiferromagnetic interactions, has a richer variety of first and second order
transitions than previously anticipated, and that such transitions are
consistent with the magnetic properties of manganites \cite{phi6af}. The mean field description of this model shows a $\varphi^{6}$ free energy \cite{antiferro}. Accordingly, one has to account for the missed phase
(s) in the phase structure of the $\left(  g\phi^{4}+h\phi^{6}\right)  $
\ field theoretic model. Moreover,   the observed large baryon asymmetry requires natural law to obey, among other principles, out-of-equilibrium dynamics. This could happen in the standard model if there was a strong first order Electroweak  phase transition \cite{baryogen}. One of the most promising techniques that results in strongly first order phase transition and also agree with the Higgs mass bounds is a $\phi^6$ Higgs self-interaction \cite{baryogen}. Accordingly, revisiting   the $\left(  g\phi^{4}+h\phi^{6}\right)  $ model to study all the possible phases is very interesting in different areas in Physics.

In this letter, we show that  the phase structure of the $\left(  g\phi
^{4}+h\phi^{6}\right)  $ theory is richer than expected in view of the newly
discovered $PT$ symmetric and non-Hermitian models  \cite{bend,bend1,bend2,bend3}.  To show this, 
we calculate the effective potential of the $\left(  g\phi^{4}+h\phi
^{6}\right)  $ model in $\left(  1+1\right)  $ dimensions up to second order
in the couplings $g$ and $h$. The obtained effective potential is investigated for the possibility of the existence of  a new phase. Indeed,  this phase is certainly existing, however, turns the theory non-Hermitian but $PT$
symmetric and thus the theory in this phase is physically acceptable. However,
to have this phase, it  leads us to the conclusion  that hermiticity can be taken as a symmetry which can
be broken or restored by  quantum corrections. \ Though hermiticity itself
is an old terminology, its consideration as a symmetry that can be broken is  new.

Now, consider the Hamiltonian density, normal-ordered\ with respect to the
mass $m$;
\begin{equation}
H=N_{m}\left(  \frac{1}{2}\left(  \left(  \triangledown\phi\right)  ^{2}%
+\pi^{2}+m^{2}\phi^{2}\right)  +\frac{g}{4!}\phi^{4}+\frac{h}{6!}\phi
^{6}\right)  . \label{h6}%
\end{equation}
 The above model is invariant
under the operation \ $H\rightarrow H^{\dagger}$. \ Let us write Eq.(\ref{h6})
in a normal-ordered\ form with respect to the mass $M=t\cdot m$, using the
following relations \cite{Coleman}:
\begin{align*}
N_{m}\phi &  =N_{M}\phi,\\
N_{m}\phi^{2} &  =N_{M}\phi^{2}+\Delta,\\
N_{m}\phi^{3} &  =N_{M}\phi^{3}+3\Delta N_{M}\phi,\\
N_{m}\phi^{4} &  =N_{M}\phi^{4}+6\Delta N_{M}\phi^{2}+3\Delta^{2},\\
N_{m}\phi^{5} &  =N_{M}\phi^{5}+10\Delta N_{M}\phi^{3}+15\Delta^{2}\phi,\\
N_{m}\phi^{6} &  =N_{M}\phi^{6}+15\Delta N_{M}\phi^{4}+45\Delta^{2}\phi
^{2}+15\Delta^{3}.
\end{align*}with
\begin{equation}
\Delta=-\frac{1}{4\pi}\ln t.
\end{equation}
Accordingly, after the application of the canonical transformation
\begin{equation}
\left(  \phi,\pi\right)  \rightarrow\left(  \psi+B,\Pi\right)  ,
\end{equation}
where $\Pi =\dot{\psi }$ and $B$ is the vacuum condensate, we can write the Hamiltonian as
\begin{equation}
H=\bar{H}_{0}+\bar{H}_{I}+\bar{H}_{1}+E, \label{hpt}%
\end{equation}
where%

\begin{align}
\bar{H}_{0}  &  =N_{M}\left(  \frac{1}{2}\left(  \Pi^{2}+\left(
\triangledown\psi\right)  ^{2}\right)  +\frac{1}{2}M^{2}\psi^{2}\right)  ,\\
\bar{H}_{I}  &  =\frac{g}{4!}N_{M}\left(  \psi^{4}+4B\psi^{3}\right) \\
&  +\frac{h}{6!}N_{M}\left(  \psi^{6}+6B\psi^{5}+\left(  15\Delta
+15B^{2}\right)  \psi^{4}+\left(  60B\Delta+20B^{3}\right)  \psi^{3}\right).
\end{align}
Also%

\begin{align*}
\bar{H}_{1}  &  =\frac{1}{2}\left(  m^{2}-M^{2}+\frac{g}{2}\left(
B^{2}+\Delta\right)  +\frac{h}{24}\left(  B^{4}+6\Delta B^{2}+3\Delta
^{2}\right)  \right)  \psi^{2}\\
&  +\left(  m^{2}+\frac{g}{6}\left(  B^{2}+3\Delta\right)  +\frac{h}%
{5!}\left(  B^{4}+10\Delta B^{2}+15\Delta^{2}\right)  \right)  B\psi,
\end{align*}
and%

\begin{align}
E  &  =\frac{1}{2}\left(  m^{2}+\frac{g}{2}\Delta\right)  B^{2}+\left(
\frac{g}{24}+\frac{h}{48}\Delta\right)  B^{4}+\frac{h}{48}h\left(  3\Delta
B^{2}+\Delta^{2}\right)  \Delta\\
&  +\frac{h}{6!}B^{6}+\frac{1}{8\pi}\left(  M^{2}-m^{2}\right)  +3g\Delta
^{2}+\frac{1}{2}m^{2}\Delta.
\end{align}
 Since $E$ serves as the generating \ functional for all the 1PI amplitudes,
it satisfies the renormalization conditions given by \cite{Peskin}%

\begin{equation}
\frac{\partial^{n}}{\partial b^{n}}E(b,t,G)=g_{n}\text{,}%
\end{equation}
where $g_{n}$ is the $\psi^{n}$ coupling. For instance,%
\begin{equation}
\frac{\partial E}{\partial B}=0\text{, \ \ \ }\frac{\partial^{2}E}{\partial
B^{2}}=M^{2}\text{,}\label{ren}%
\end{equation}
These conditions enforces $\bar{H}_{1}$ to be zero and thus
\begin{align*}
\frac{1}{2}\left(  m^{2}-M^{2}+\frac{g}{2}\left(  B^{2}+\Delta\right)
+\frac{h}{24}\left(  B^{4}+6\Delta B^{2}+3\Delta^{2}\right)  \right)   &
=0,\\
\left(  m^{2}+\frac{g}{2}\left(  B^{2}+3\Delta\right)  +\frac{h}{5!}\left(
B^{4}++10\Delta B^{2}+15\Delta^{2}\right)  \right)  B &  =0.
\end{align*}
 The use of the dimensionless parameters \ $t=\frac{M^{2}}{m^{2}}$,
$G=\frac{g}{2\pi m^{2}}$, $H=\frac{h}{\left(  4\pi m\right)  ^{2}}$ and
$b^{2}=4\pi B^{2}$, leads to the following results
\begin{align}
\left(  6b^{2}-6\ln t\right)  \frac{G}{4!}+\left(  15b^{4}-90b^{2}\ln
t+45\ln^{2}t\right)  \allowbreak\frac{H}{6!}+1 &  =t,\nonumber\\
2b\left(  \left(  2b^{2}-6\ln t\right)  \frac{G}{4!}+\left(  3b^{4}-30b^{2}\ln
t+45\ln^{2}t\right)  \allowbreak\frac{H}{6!}+1\right) \allowbreak &
=0,\label{orphi6}%
\end{align}%
\begin{equation}
E=\frac{1}{4\pi}m^{2}\left(
\begin{array}
[c]{c}%
\left(  b^{2}+\frac{1}{2}\left(  t-1\right)  -\ln t\right)  +\left(
b^{4}-6b^{2}\ln t+3\ln^{2}t\right)  \frac{G}{4!}\\
+\frac{H}{6!}\left(  b^{6}-15b^{4}\ln t+45b^{2}\ln^{2}t-15\ln^{3}t\right).
\end{array}
\right)  \label{En1st}%
\end{equation}
For some specific values of $G$ and $H$, one solves Eq.(\ref{orphi6}) to get
the values of $b$ and $t$. Thus, as $t$ chosen to be greater than zero, the
solutions determine the parameters at the minima of the energy density.

\begin{figure}[ht]
	\begin{center}
		\includegraphics{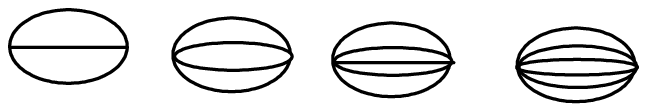}
	\end{center}
	\caption{The non-cactus vacuum diagrams (up to $g^{2}$ and $h^{2}$) contributing to  the effective potential of
$(\frac{g}{4!}\phi^{4}+\frac{h}{6!}\phi^{6})$ theory.}
	\label{feyngh2}
\end{figure}

The normal ordered effective potential obtained above agrees with GEP results \cite{gep6} which in turn accounts
not only for the leading order diagrams but also for all the non-cactus
diagrams \cite{wen-Fa,changcac}. Thus, to go to higher orders we include only
non-cactus diagrams. Up to \ second order in the couplings, we have the non-cactus diagrams shown in
Fig.\ref{feyngh2}. The general form of these diagrams contributions to the effective
potential is
\[
\frac{1}{-i}%
{\displaystyle\prod\limits_{l}}
\frac{i\Gamma\left(  n-%
{\displaystyle\sum\limits_{l=1}^{L}}
\frac{ld}{2}\right)  }{\left(  4\pi\right)  ^{\frac{d}{2}}\Gamma\left(  n-%
{\displaystyle\sum\limits_{l=1}^{L-1}}
\frac{ld}{2}\right)  S}V\left(  g,h,B\right)  I,
\]
where  $V(b,g,h)$ represents vertices of the diagram, $S$ is the symmetry factor ($S=2(L-1)!$),
$L$ is the number of loops in the diagram and $I$ is the integral over the Feynman parameters. For the last diagram (5-loop diagram) $I$ has the form%
\[
I=%
{\displaystyle\int\limits_{0}^{1}}
d^{L}x\frac{\delta\left(
{\displaystyle\sum\limits_{i=1}^{6}}
x_{i}-1\right)  }{\left[  \left(  x_{2}x_{3}x_{4}x_{5}x_{6}\right)  +\left(
x_{1}x_{3}x_{4}x_{5}x_{6}\right)  +......\left(  x_{1}x_{2}x_{3}x_{4}%
x_{5}\right)  \right]
{\displaystyle\sum\limits_{i=1}^{6}}
x_{i}}\text{.}%
\]
The integral was computed numerically using Monte Carlo method when a
straightforward integration was not possible.

We obtained the following form for the effective potential up to second
order in the couplings:%
\[
E=\left(
\begin{array}
[c]{c}%
\left(  b^{2}+\frac{1}{2}\left(  t-1\right)  -\ln t\right)  +\left(
b^{4}-6b^{2}\ln t+3\ln^{2}t\right)  \frac{G}{4!}\\
+\frac{H}{6!}\left(  b^{6}-15b^{4}\ln t+45b^{2}\ln^{2}t-15\ln^{3}t\right) \\
-8.\,\allowbreak297\,9\times10^{-4}\frac{H^{2}\allowbreak}{t}-1.\,\allowbreak
005\,6\times10^{-3}\frac{H^{2}}{t}\allowbreak b^{2}\\
\frac{-8.\,\allowbreak764\,6\times10^{-2}}{2t}\allowbreak\left(  G+Hb^{2}-H\ln
t\right)  ^{2}\allowbreak-5.\,\allowbreak425\,7\times10^{-3}b^{2}%
\allowbreak\frac{1}{t}\left(  Hb^{2}+3G-3H\ln t\right)  ^{2}\allowbreak
\end{array}
\right)  ,
\]
subjected to the stability condition $\frac{\partial E}{\partial b}=0$. As
usual, we use the renormalization conditions to get the renormalized couplings.
For instance
\[
M^{2}=\frac{\partial^{2}E}{\partial B^{2}}\text{, } \ \ \ g_{r}=\frac{\partial^{4}%
E}{\partial B^{4}}\ \ \text{ and }\ \ \ \ h_{r}=\frac{\partial^{6}E}{\partial B^{6}%
}\text{.}%
\]
Our form for the effective potential implemented a renormalization group
invariance of the bare parameters on the scale $t$. However, to make sure that
the effective theory and the original one are totally equivalent as
$t\rightarrow1$, we introduced a new renormalization group equation. Besides
the scale invariance of the bare parameters $m_{0}$, $g_{0}$ and $h_{0}$, we
added the scale invariance of the bare vacuum energy (it is certainly zero,
but we fix this zero to be scale invariant). In fact, normal ordering do this
automatically as can be seen from Eqs.(\ref{hpt}),(\ref{orphi6}) and (\ref{En1st}
), as $t\rightarrow1$, the effective Hamiltonian (Eq.(\ref{hpt})) tends to the
original Hamiltonian in Eq.(\ref{h6}). For higher orders, however, without the introduction of
the new renormalization group invariance, we can not get this
equivalence and thus both theories are inequivalent.

Our result for the effective potential verifies all the known results for the
the $\left(  g\phi^{4}+h\phi^{6}\right)  $ \ field theoretic model, second
order phase transition for $g>0$ and first order phase transition for $g<0$.
Moreover, a new phase with negative condensate squared has been investigated
for which the theory is non-Hermitian but $PT$ symmetric. The unbroken $PT$ symmetry assure
the physical acceptability of the theory in this phase. 

The negative sign of the condensate squared is technical and not conceptual because it is related to
the expected negative norm of the theory in this phase. This problem can be
remedied by calculating the $C$ operator of this theory and the correct inner
product \cite{coper,coper1}
\[
\langle A|B\rangle_{CPT}=(CPT|A\rangle)^{T}|B\rangle,
\]
to be used. In fact, this has been done for another model for which $PT$ symmetric non-Hermitian formulation saved its
validity, namely, the Lee model \ which was introduced in the 1950s as an elementary
quantum field theory in which mass, wave function, and charge renormalization
could be carried out exactly. In early studies of this model it was found that
there is a critical value of $g^{2}$, the square of the renormalized coupling
constant, above which $g_{0}^{2}$, the square of the unrenormalized coupling
constant, is negative. Thus, for $g^{2}$ larger than this critical value, the
Hamiltonian of the Lee model becomes non-Hermitian. It was also discovered
that in this non-Hermitian regime a new state appears whose norm is negative.
This state is called a ghost state. It has always been assumed that in this
ghost regime the Lee model is an unacceptable quantum theory because unitarity
appears to be violated. However, in this regime while the Hamiltonian is not
Hermitian, it does possess PT symmetry. \ Again, the construction of an inner
product via the construction of a linear operator $C$ saves the theory from
physical unacceptability \cite{lee}.
However, this calculation for the model we are studying is out of the scope of this letter and
it naturally becomes a topic of our future work to overcome the sign problem
of the condensate squared.

\begin{figure}
	\begin{center}
		\includegraphics{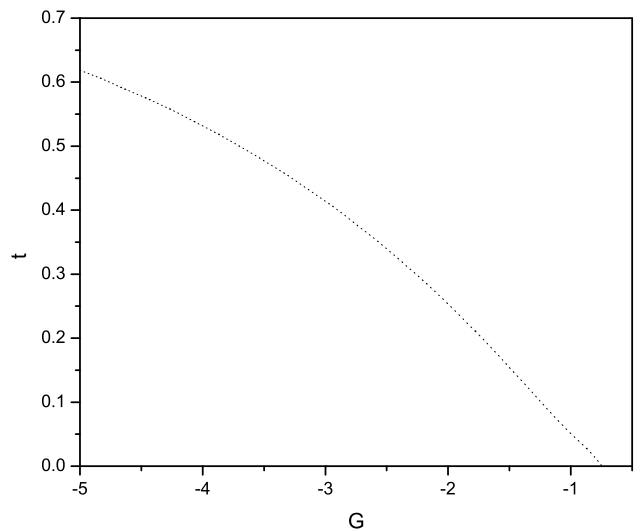}
	\end{center}
	\caption{The reciprocal of the 1PI two-point function versus the coupling $G$ for second order in the perturbation series and for $H=0.1$ for the $PT$ symmetric and non-Hermitian phase. At the critical coupling, we realize that the mass parameter vanishes and thus the phase transition is of the second order type.}
	\label{tghp1}
\end{figure}

\begin{figure}
	\begin{center}
		\includegraphics{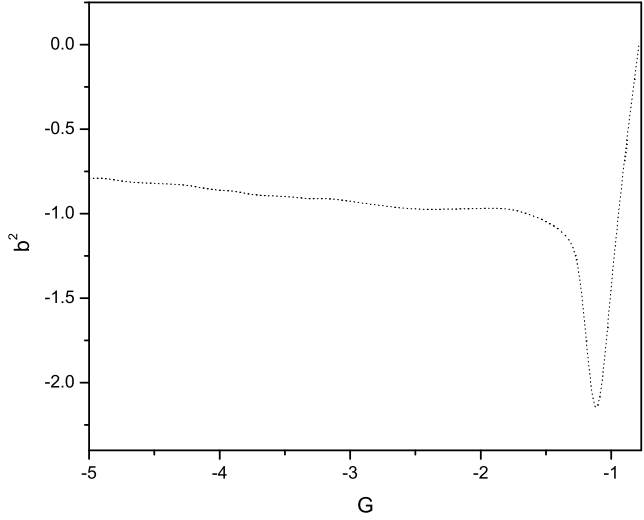}
	\end{center}
	\caption{The vacuum condensate squared versus the coupling $G$ for the second order in the perturbation series and for $H=0.1$ for the $PT$ symmetric and non-Hermitian phase. Again, the diagram assures the second order phase transition though for negative $G$ values.}
	\label{bghp1}
\end{figure}

\begin{figure}
	\begin{center}
		\includegraphics{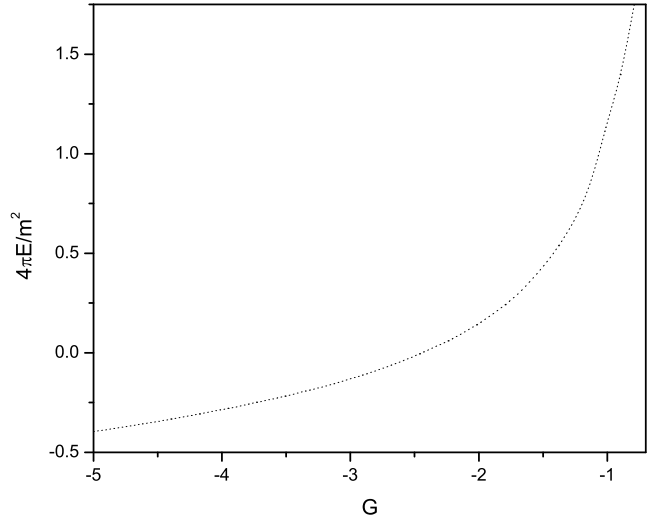}
	\end{center}
	\caption{The  vacuum energy $E$ as a function of the coupling $G$ for the second order in the perturbation series and for $H=0.1$ for the $PT$ symmetric and non-Hermitian phase.}
	\label{vacenghp1}
\end{figure}

The parameters of the new phase ($PT$ symmetric) as well as the vacuum energy of this phase for the ($g\phi^{4}+h\phi^{6})_{1+1}$ model are shown in Figs. \ref{tghp1}, \ref{bghp1} and \ref{vacenghp1}, respectively ( for $H=0.1$). As we can see from the mass parameter and  the vacuum condensate  diagrams, the phase transition is of second order type. 

Since $b^2$ represents the number of condensed Bosons \cite{thermo}, its negative sign is an indication of antiparticles. However, the first order phase existing also for negative $g$ (not shown in the diagrams) has a bigger vacuum condensate which is real and thus represent matter phase. Accordingly, this model may offer a scenario for the matter-antimatter asymmetry in the universe.

To account for the reliability of the order of calculations we carried out, we made sure that  the effective potential passed   tests for the known features  like the existence of second order phase transition for $g$ positive and the existence of a first order phase transition for $g$ negative. This agrees well with a previous numerical calculations \cite{montecarlo}.  Also, in the region of interest, even mean field calculations suffices to describe the theory \cite{wilczk}. Moreover, the perturbative characteristics of the model used have been defended in Ref.\cite {orpap}.

To conclude, we calculated the effective potential of the Hermitian $\left(  g\phi^{4}+h\phi^{6}\right)$ field theory up to second order in the couplings $g$ and $h$. Also, in our calculation of the effective potential, we introduced a new renormalization group equation, namely, the invariance of bare vacuum energy under change of scale. Without this renormalization group equation, higher orders corrections to the effective potential spoils out the equivalence between the effective theory and the original one.   We find a new phase for the Hermitian $\left(  g\phi^{4}+h\phi^{6}\right)$ field theory. This phase turns the theory non-Hermitian but $PT$ symmetric and thus it is physically acceptable. This phase may resemble the para-magnetic to anti-Ferro magnetic phase transition in statistical systems. We interpreted this phase as a phase of antimatter and it is less dense than the first order matter phase. Accordingly, this model with the new phase resembles matter-antimatter asymmetry. 

\section*{Acknowledgment}

\label{ack} The author would like to thank Dr. S.A. Elwakil for his support
and kind help. Also, deep thanks to Dr. C.R. Ji, from North Carolina State
University, for his direction to my attention to the critical phenomena in QFT
while he was \ supervising my Ph.D.


\newpage

\end{document}